\documentstyle[PASJadd]{PASJ95}
\draft
\newcommand{\vol}{52}
\newcommand{\no}{6}
\newcommand{\lett}{}
\newcommand{\spage}{}
\newcommand{\rdate}{2000 June 16}
\newcommand{\adate}{2000 August 16}
\begin{document}
\title{Resistive Magnetohydrodynamics of Jet Formation and
       Magnetically Driven Accretion}

\author{Takuhito {\sc Kuwabara}\\
{\it Graduate School of Science and Technology, Chiba University,
Inage-ku, Chiba 263-8522}\\
{\it E-mail(TK): takuhito@c.chiba-u.ac.jp}\\[6pt]
Kazunari {\sc Shibata}\\
{\it Kwasan and Hida Observatories, Kyoto University, Yamashina,
Kyoto 607-8471}\\
Takahiro {\sc Kudoh}\\
{\it National Astronomical Observatory, Mitaka, Tokyo 181-8588}\\
and\\
Ryoji {\sc Matsumoto}\\
{\it Department of Physics, Faculty of Science, Chiba University,
Inage-ku, Chiba 263-8522}}

\abst{We carried out 2.5-dimensional resistive magnetohydrodynamic
simulations to study the effects of magnetic diffusivity on
magnetically driven mass accretion and jet formation. 
The initial state is
a constant angular-momentum torus threaded by large-scale
vertical magnetic fields. Since the angular momentum of the torus is 
extracted due to 
magnetic braking, the torus medium falls toward the central region.
The infalling matter twists the large-scale magnetic fields and 
drives bipolar jets.
We found that (1) when the normalized magnetic diffusivity, 
$\bar{\eta}\equiv \eta/(r_0 V_{\rm K0})$, 
where $V_{\rm K0}$ is the Keplerian rotation speed at a reference radius
$r=r_{\rm 0}$, is small
($\bar{\eta}\leq 10^{-3}$), mass accretion and jet formation take place 
intermittently;
(2) when $10^{-3}\leq\bar{\eta}\leq 10^{-2}$, 
the system evolves toward a quasi-steady state; and
(3) when $\bar{\eta}\geq 10^{-2}$ the accretion/mass outflow rate decreases
with $\bar{\eta}$ and approaches 0.
The results of these simulations indicate
that in the center of a galaxy which has a super-massive 
($\sim 10^9$ $M_{\odot}$) black hole, a massive ($\sim 10^8$ $M_{\odot}$) 
gas torus
and magnetic braking provide a mass accretion
rate which is sufficient to explain the activity of AGNs when 
$\bar{\eta}\leq 5\times 10^{-2}$.
} 

\kword{accretion, accretion disks --- galaxies: active --- galaxies: nuclei 
       --- galaxies: jets --- MHD} 

\maketitle
\thispagestyle{headings}

\section
{Introduction}

Magnetically driven outflows from accretion disks are the most promising
models of the acceleration and collimation of jets/winds in AGNs and
in star-forming regions. By assuming steady axisymmetric cold outflow,
it has been shown that when the magnetic field lines make angles of less than
$60^{\rm o}$ from the equatorial plane, a magneto-centrifugally 
driven outflow of matter
emanates from the surface of the disk 
(Blandford, Payne 1982; Pudritz, Norman 1983; Lovelace et al. 1986; 
 Shu et al. 1994a,b).
Recently, by including gas pressure, Kudoh and Shibata (1997) obtained
a steady solution along a magnetic field line 
and showed that when the strength of the poloidal magnetic field, $B_{\rm p0}$,
is much smaller than the toroidal component, $B_{\varphi 0}$, 
the mass outflow rate, $\dot{M}_{\rm jet}$, increases with $B_{\rm p0}$
and $\dot{M}_{\rm jet}$ approaches a constant value 
when $B_{\rm p0}>B_{\varphi 0}$.

Several authors have reported the results of 
axisymmetric two-dimensional
magnetohydrodynamic (MHD) simulations of jet formation 
by fixing the boundary conditions at 
the disk surface without including the effects of magnetic braking on the disk
(e.g., Ustyugova et al. 1995; Meier et al. 1997; Ouyed et al. 1997). 
The surface conditions of the accretion disk, however, should be determined
self-consistently. 
Uchida and Shibata (1985) as well as Shibata and Uchida (1986) carried out
nonlinear, two-dimensional MHD simulations of jet
formation from accretion disks by including the effects of a back reaction of
jet formation on disks.
They showed that jets/winds are accelerated along
the magnetic field lines twisted by the rotation of the disk.
They called this mechanism a ``sweeping magnetic twist mechanism''.
The numerical results by Uchida and Shibata (1985) as well as
Shibata and Uchida (1986) have been confirmed by Stone and Norman (1994)
by using the ZEUS code. Stone and Norman (1994) have also
discussed the relation between magnetic braking and
magnetorotational instability (Balbus, Hawley 1991).

The effects of magnetic extraction of angular momentum (magnetic braking)
on the disk become more evident when a
geometrically thick disk (or torus) is considered. 
Matsumoto et al. (1996) carried out 2D MHD simulations of a torus
threaded by poloidal magnetic fields and showed that the surface layer of 
the torus accretes faster than the equatorial region, like an avalanche,
because magnetic braking most efficiently extracts angular momentum
from that layer. 
Kudoh, Matsumoto, and Shibata (1998) confirmed the numerical results 
by employing a newly developed 
CIP-MOCCT code which uses the CIP method (Yabe, Aoki 1991) for 
hydrodynamic part and the MOCCT scheme 
(Stone, Norman 1992) to solve the induction equations and to evaluate
the Lorentz force terms. 
They have proposed that the ejection mechanism of non-steady jets found
in the 2.5-dimensional simulations can be understood using the steady
state theory even when the back reaction of
the jet on the disk is included self-consistently.
Matsumoto and Shibata (1997) extended the 2D model to
3D by taking into account the non-axisymmetric effects, and
have shown that the avalanche breaks up into several spiral channels.
When the torus is threaded by large-scale poloidal magnetic field lines,
accretion proceeds along these spiral channels. Since the channel
flow bundles the large-scale magnetic field lines, a spiral structure
also appears in the jet. 

Most of the nonsteady models of 
magnetohydrodynamic jet formation from an accretion disk have
assumed ideal MHD. Recently, several authors included resistivity to
study protostellar jets. Hayashi et al. (1996) carried out 2D 
resistive MHD simulations of the interaction of the dipole 
magnetic field of the protostar and its surrounding disk and
have shown that magnetic reconnection takes place in the current
sheet created inside the expanding magnetic loops. 
They successfully explained both the acceleration of optical jets
and the X-ray flares in protostars observed by the ASCA satellite
(Koyama et al. 1996). 
Similar resistive MHD simulations of protostellar
jets have been carried out by Miller and Stone (1997), 
Grosso et al. (1997), and Goodson et al. (1997, 1999).

Hirose et al. (1997) investigated the interaction of 
the stellar magnetosphere originated in its dipole magnetic field
and the interstellar magnetic field carried with the infalling gas. 
They carried out 2D
resistive MHD simulations and showed that magnetic reconnection takes place
by an interaction of these magnetic fields. 
They showed that matter accretes along the reconnected magnetospheric
field and that the magneto-centrifugal force accelerates 
the reconnection-driven jet.

Resistivity can also play important roles
when the accretion disk is threaded by open magnetic fields.
Lubow, Papaloizou, and Pringle (1994), Ogilvie (1997), Kudoh and Shibata (1997)
tried to obtain steady solutions of magnetically driven winds/jets
including the accretion disk. 
But, in reality,
when resistivity is not included,
the accretion disk may be highly nonsteady due to surface 
avalanching flow (Matsumoto et al. 1996; Kudoh et al. 1998) and
due to magnetorotational instability inside the disk (Balbus,
Hawley 1991; Hawley, Balbus 1992). 
For dealing with a steady state, the inclusion of resistivity is essential.
Kaburaki (1987) analytically obtained a steady-state solution
for a disk of finite thickness by using resistive MHD equations.
It has been shown that in 3D,
magneto-rotational instability generates turbulence inside the disk
(Hawley et al. 1995; Matsumoto, Tajima 1995;
Brandenburg et al. 1995). This turbulence generates an
effective magnetic diffusivity which can suppress the growth of
the magnetorotational instability. It is possible that an accretion
disk can be in a marginally stable state in which the magnetic turbulence
is maintained at a marginal level over which turbulent magnetic diffusivity 
kills the growth of the magnetorotational instability
(Matsumoto, Tajima 1995).

In weakly ionized disks, since the Spitzer-type resistivity itself
becomes large, it can affect the growth of the magnetorotational
instability (Sano et al 1998). In this work, we introduced resistivity
to simulate the effects of either the turbulent magnetic diffusivity
or the resistivity in very weakly ionized disks on the formation
of jets. 
The effects of resistivity are that magnetic field lines do not
rotate with the same angular speed as the disk matter,
and thus it suppresses the injection of 
magnetic helicity (magnetic twists) and the magneto-centrifugal acceleration. 
We would like to study the dependence of
the mass accretion rate, mass outflow rate, and jet speed on resistivity
(or turbulent diffusivity) by 2D axisymmetric resistive MHD simulations.

Observations by the Hubble Space Telescope (Jaffe et al. 1996) indicate that 
a several hundred parsec scale rotating gas torus exists in 
active galactic nuclei.
When the torus is threaded by large-scale magnetic field lines,
twist injection from the disk drives outflows. 
Since the magneto-rotational instability drives magnetic turbulence
in the disk,
the turbulent magnetic diffusivity can be important in such disks.

In section 2, we present physical assumptions and numerical methods.
Numerical results are given in section 3. Section 4 is devoted to a summary and
discussion.

\section
{Physical Assumptions and Numerical Methods}

We assume that at the initial state a differentially rotating 
polytropic torus surrounding a central gravitating object is 
threaded by a uniform vertical magnetic field.
In active galactic nuclei, the torus corresponds to a molecular
torus rotating around the central super-massive black hole.
The large-scale poloidal magnetic field threading the torus can
be produced by interstellar magnetic fields swept into the
nuclear region of the galaxy with the gas constructing the torus.

We assume axisymmetry and neglect the effects of self-gravity 
and cooling. Although the molecular torus in AGNs is in low ionization
state, we assume that ambipolar diffusion is negligible and that
the gas torus can be treated by using resistive, single fluid MHD 
equations. We discuss the validity of these assumptions further 
in section 4.

We use cylindrical coordinates ($r$, $\varphi$, $z$) and assume that
$z$-direction is parallel to the rotation axis. A schematic
picture of the simulation model is shown in figure 1. The basic 
equations are:
\begin{equation}
   \frac{\partial\rho}{\partial t}+\mbox{\boldmath $\nabla$}\cdot 
(\rho\mbox{\boldmath$v$})=0,
\end{equation}
\begin{eqnarray}
   \frac{\partial}{\partial t}(\rho v_{r})
   +\mbox{\boldmath $\nabla$}\cdot (\rho v_{r}\mbox{\boldmath$v$})
   -\frac{\rho v^2_{\varphi}}{r}+
    \frac{\partial P}{\partial r} & \nonumber \\
   &\displaystyle
    \mbox{\hspace{-3cm}}-\frac{1}{4\pi}[(\mbox{\boldmath $\nabla$}
                                    \times\mbox{\boldmath$B$})
                                    \times\mbox{\boldmath$B$}]_{r}
   +\rho\frac{\partial\psi}{\partial r}=0, 
\end{eqnarray}
\begin{eqnarray}
   \frac{\partial}{\partial t}(\rho v_{z})
   +\mbox{\boldmath $\nabla$}\cdot (\rho v_{z}\mbox{\boldmath$v$})
   +\frac{\partial P}{\partial z} & \nonumber \\
   &\displaystyle
    \mbox{\hspace{-2cm}}-\frac{1}{4\pi}[(\mbox{\boldmath$\nabla$}
                                        \times\mbox{\boldmath$B$})
                                        \times\mbox{\boldmath$B$}]_{z} 
   +\rho\frac{\partial\psi}{\partial z}=0,
\end{eqnarray}
\begin{equation}
   \frac{\partial}{\partial t}(r\rho v_{\varphi})
   +\mbox{\boldmath$\nabla$}\cdot (r\rho v_{\varphi}\mbox{\boldmath$v$}) 
   -\frac{1}{4\pi}\mbox{\boldmath$\nabla$}\cdot
   (rB_{\varphi}\mbox{\boldmath$B$})=0,
\end{equation}
\begin{equation}
   \frac{\partial\mbox{\boldmath$B$}}{\partial t}
   -\mbox{\boldmath$\nabla$}\times(\mbox{\boldmath$v\times B$})
   -\eta\mbox{\boldmath$\nabla$}^2\mbox{\boldmath$B$}=0,
\end{equation}
\begin{equation}
   \left(\frac{\partial}{\partial t}+\mbox{\boldmath$v\cdot\nabla$}\right)
   \left(\rho e \right)+P\mbox{\boldmath$\nabla$} \cdot\mbox{\boldmath$v$}
   = \eta  \left(\frac{\mbox{\boldmath$\nabla\times B$}}{4\pi}\right)^2. 
\end{equation}
In these equations, $\rho$, $P$, and $\gamma$ are the density, pressure,
and specific heat ratio, respectively;
{\boldmath$v$} the velocity of the gas;
$e$ is the internal energy of the gas $e=P/[(\gamma -1)\rho]$;
{\boldmath$B$} the magnetic field and $\psi$ is the gravitational potential,
which we assume to be
\begin{equation}
   \psi =-\frac{GM}{(r^2+z^2)^{1/2}},
\end{equation}
where
$G$ is the gravitational constant and $M$ is the mass of the central object. 
The resistivity, $\eta$, is assumed to be uniform.
We normalize the physical quantities with those 
at ($r$, $z$) = ($r_0$, $0$) where the initial density is maximum.
In this normalization, 
we have two non-dimensional parameters:
\begin{equation}
   E_{\rm th}=\frac{V^2_{\rm s0}}{\gamma V^2_{\rm K0}}
         \sim 10^{-2}\left(\frac{T}{10^4\mbox{ } {\rm K}}\right)
              \left(\frac{M}{10^8\mbox{ }M_{\odot}}\right)^{-1}
              \left(\frac{r}{100\mbox{ }{\rm pc}}\right) ,
\end{equation}
\begin{equation}
   E_{\rm mg}=\frac{V^2_{\rm A0}}{V^2_{\rm K0}}
             \sim 10^{-4}\left(\frac{B}{10^{-3}\mbox{ }{\rm G}}\right)
             \left(\frac{\rho}{10^{17}\mbox{ }{\rm g\ cm^{-3}}}\right)^{-1}
             \left(\frac{M}{10^8\mbox{ }M_{\odot}}\right)^{-1}
             \left(\frac{r}{100\mbox{ }{\rm pc}}\right) ,
\end{equation}
where $V_{\rm s0}=\sqrt{\gamma P_0/\rho_0}$ and $V_{\rm K0}=\sqrt{GM/r_0}$
are the sound speed and Keplerian rotation speed at ($r$, $z$)
= ($r_0$, $0$), respectively.
Here, $E_{\rm th}$ is the ratio of thermal energy to gravitational energy
and $E_{\rm mg}$ is the ratio of magnetic energy to gravitational energy.
The Alfv$\rm{\acute{e}n}$ speed is defined as 
$V_{\rm A0}=B_0/(4\pi\rho_0)^{1/2}$, where
$\rho_0$ and $B_0$ are the initial density and magnetic field strength 
at ($r$, $z$) = ($r_0$, $0$).

Exact equilibrium solutions of a torus can be obtained 
under the following simplifying 
assumptions for the distributions of angular momentum and pressure of
the rotating torus (Matsumoto et al. 1996):
\begin{equation}
   L=L_0r^a
\end{equation} 
and
\begin{equation}
   P=K\rho^{1+1/n}.
\end{equation}
The density distribution of the torus can be determined by
\begin{eqnarray}
   -\frac{GM}{(r^2+z^2)^{1/2}}+\frac{1}{2(1-a)}L_0^2r^{2a-2}
     +(n+1)\frac{P}{\rho}={\rm constant}.
\end{eqnarray} 
The mass distribution outside the torus is assumed to be that of 
the isothermal nonrotating high-temperature halo surrounding 
the black hole,
\begin{equation}
   \rho=\rho_{\rm h} {\rm exp}\left[\alpha\left(\frac{r_0}{\sqrt{r^2+z^2}}
   -1\right)\right],
\end{equation}
where $\alpha=\gamma V^2_{\rm K0}/V^2_{\rm sc}$. Here $V_{\rm sc}$ 
and $\rho_{\rm h}$
are the sound velocity and density in the halo
at ($r$, $z$) = ($0$, $r_0$), respectively.
We assumed that $\alpha =1.0$ and $\rho_{\rm h}/\rho_0=10^{-3}$ throughout
this work.
We also use normalizations $r_0=V_{\rm K0}=\rho_0=1.0$.

The numerical method which we used was a modified Lax-Wendroff scheme with
artificial viscosity for 2.5-dimensional MHD problems in cylindrical
geometry (Rubin, Bustein 1967; Richtmeyer, Morton 1967). 
The code was originally developed by Shibata (1983) and 
has been extended by Matsumoto et al. (1996) and Hayashi et al. (1996).

The boundary conditions were 
as follows: At the equatorial plane ($z=0$), we assume a boundary condition 
that is symmetric for $\rho$, $P$, $v_{r}$, $v_{\varphi}$, 
and $B_{z}$ but
antisymmetric for $v_{z}$, $B_{r}$, and $B_{\varphi}$. 
On the rotation axis ($r=0$), 
$\rho$, $P$, $v_{z}$, and $B_{z}$ are symmetric, while $v_{r}$, 
$v_{\varphi}$,
$B_{r}$, and $B_{\varphi}$ are zero. The side surface $r=R_{\rm max}$, 
and the top surface $z=Z_{\rm max}$, are free boundaries at which the mass 
as well as waves can go through freely, and we set 
$\partial\delta Q/\partial z =0$ on $z=Z_{\rm max}$ and 
$\partial\delta Q/\partial r=0$ on $r=R_{\rm max}$, where $Q$ is one of the 
above variables, and $\delta Q\equiv Q(r, z, t+\delta t)-Q(r, z, t)$. 
The region 
around $r=z=0$ is treated by softening the gravitational potential as
$\psi =-GM/(r^2+z^2+\epsilon^2)^{1/2}$, where $\epsilon =0.2\mbox{ }r_0$.
We set wave absorbing condition in the region where 
$\sqrt{r^2+z^2} < 0.1\mbox{ }r_{\rm 0}$.
In this region we correct the density, pressure, and 
radial component of magnetic field $B_{r}$ by
$Q^{'}=Q-\alpha^{'}(Q-Q_0)$, where $Q_{\rm 0}$ denotes the initial value 
and $\alpha^{'}=0.5\{1-{\rm tanh}[400(\sqrt{r^2+z^2}/r_0-0.075)]\}$.
Furthermore, we correct the toroidal velocity by
$v^{'}_{\varphi}=v_{\varphi}-0.01\alpha^{'}v_{\varphi}$.

The size of the simulation box was $0\leq r<5.1\mbox{ }r_0$ 
and $0\leq z<13.4\mbox{ }r_0$. 
The minimum grid size was $0.01\mbox{ }r_0$.
The number of grid points was
$201 \times 256$ for the models presented in this paper. 
The grid spacing was stretched  
when $r/r_0>1$ or $z/r_0>1$.

\section
{Numerical results}

The simulation models discussed in this work are shown in table~1.
For all models, we took $a=0$ $(L={\rm constant})$, $n=3$, $\gamma=5/3$,
$E_{\rm th}=0.05$, $E_{\rm mg}=5.0\times 10^{-4}$ and 
$\rho_{\rm h}/\rho_0=10^{-3}$.
The initial magnetic field was assumed to be uniform and parallel to the 
$z$-axis.
The normalized resistivity $\bar{\eta}$ is defined by 
$\bar{\eta}=\eta /(r_{\rm 0}V_{\rm K0})$.

Figure 2 shows the time evolution of a typical resistive model (model 5),
where $\bar{\eta}=0.0125$.
The magnetic Reynolds number at the density maximum of the torus 
defined by 
\begin{equation}
R_{\rm m0}\equiv r_0V_{\rm A0}/\eta 
\end{equation}
is $R_{\rm m0} =1.8$.
The magnetic Reynolds number near the surface of the torus is 
\begin{equation}
R_{\rm mh}\simeq r_0V_{\rm Ah}/\eta =r_0(V_{\rm A0}/\eta) 
(V_{\rm Ah}/V_{\rm A0})
=r_0(V_{\rm A0}/\eta) (\rho_0/\rho_{\rm h})^{1/2}\sim 57.
\end{equation}
Here, $V_{\rm Ah}$ is the Alfv\'{e}n speed at ($r$, $z$) = ($1.0$, $0.75$).
The poloidal magnetic field lines depicted by integrating the equation
of magnetic lines of force
(bottom panels) show that the surface layer
of the torus falls faster than the equatorial plane due to magnetic
braking. 
For comparison, we show in figure 3 
the results of non-resistive model (model 1). 
The bottom panels show the isocontours of $rA_{\varphi}$ 
($A_{\varphi}$ is the vector potential)
which depict poloidal magnetic field lines when $\bar{\eta}=0$.
After one rotation
period ($t=2\pi$), the magnetic field lines near 
the surface are already convected to the central gravitating object
in the non-resistive model (figure 3). 
In the resistive model (figure 2),
however, this surface avalanche (Matsumoto et al. 1996) is not so evident as
that in non-resistive model;
the magnetic field lines are
only slightly deformed. 

The speed of the jet in the early stage ($t<9$) 
of the resistive model is slower than that in the non-resistive model.
However, as the magnetic field lines are deformed and
the angle
from the equatorial plane decreases ($t=18.5$), the jet speed increases
and approaches the Keplerian rotation speed of the torus.

In the non-resistive model (figure 3), the jet-forming surface area of 
the torus is narrower than that of the resistive model
because some part of the surface layer moves outward by 
gaining angular momentum, thus preventing the infall of the outer disk material.
Inside the torus, strong twists are accumulated, as shown in the isocontours
of $B_{\varphi}$ in both the non-resistive model and the resistive model.
The torus expands in the vertical direction
due to magnetic pressure produced 
by the accumulated toroidal magnetic fields.

In the resistive model (figure 2),  
the magnetic field lines inside the torus
approach straight lines (see the bottom panels of figure 2).
The torus is deformed into
a disk-like shape and accretion proceeds inside the torus by
magnetically losing angular momentum.
Figure 4 shows the numerical results for a highly resistive model 
(model 7; $\bar{\rm \eta}=0.05$, $R_{\rm m0}=0.4$, $R_{\rm mh}\sim 13$).
In this model, since $R_{\rm m0}<1$, the torus is stable for 
the magnetorotational
instability (Sano et al. 1998). These results
indicate that the magnetic field lines are deformed only slightly 
and that magnetic twists are not accumulated inside the torus.
Only weak outflow with velocity $v\sim 0.2V_{\rm K0}$ appears in this model.

Figure 5--7 show 3D pictures of the magnetic field lines and isosurface of
density for models 5, 1, and 7, respectively. As the resistivity increases,
the angle which the large-scale magnetic field lines make with the
vertical axis decreases. The magnetic twist also decreases
with the resistivity. The density isosurfaces evidently show that bipolar
outflow is produced in model 5 and model 1.

Figure 8 shows the time variation of the temperature distribution 
(gray scale) and 
$r A_{\rm \varphi}$ ($A_{\rm \varphi}$ is vector potential), 
which  depicts 
magnetic field lines and velocity vectors for models 5, 1, and 7.
Figure 8a (upper three panels) is for model 5. 
In this mildly diffusive model, mass accretion and mass outflow
take place continuously. Figure 8b (middle three panels) is for model 1. 
In this non-diffusive model,
the mass accretion and mass outflow take place episodically.
At $t=5.96$, the first avalanching mass accretion 
and mass outflow take place. 
Subsequently, the accretion rate decreases 
because some parts of the disk material
obtain angular momentum and prevent the outer region from infalling. 
Around $t=15.9$, surface mass accretion occurs again.
Figure 8c (bottom three panels) is for model 7.
Almost no mass accretion and mass outflow take place.

Figures 9a and b show the mass outflow rate for various models. 
The mass outflow rate is defined as 
\begin{equation}
\dot{M}_{\rm jet}=2\times\int_0^{3.0}2\pi r \rho V_{z} dr
\end{equation}
at $z=3.1\mbox{ }r_{\rm 0}$.
When $\bar{\eta}\sim 0.0125$, 
it increases monotonically with time 
and approaches a quasi-steady value (model 5, model 6).
When $\bar{\eta} <0.0125$, however, the mass outflow rate has several peaks
(model 1, model 2, model 3, model 4). In these models jet production 
takes place episodically. When $\bar{\eta} > 0.0125$ (model 7) the mass outflow
rate approaches zero. 
Figures 10a and b show the time variation of the mass accretion rate,
defined as 
\begin{equation}
\dot{M}_{\rm acc}=2\times\int_{0}^{0.4} 2\pi r \rho v_{r} dz
\end{equation}
at $r=0.3\mbox{ }r_{\rm 0}$.
When $\bar{\eta}$ takes a value between $0$ and $0.01$, 
the time averaged-accretion rate when $t\gg 0$ takes almost the same value 
independent of $\bar{\eta}$, although the accretion rate increases 
more rapidly with time when $\bar{\eta}\sim 0$.
When $0.01<\bar{\eta}<0.05$, the accretion rate approaches
a quasi-steady value, but the peak accretion rate is smaller than in
models with $\bar{\eta} <0.01$. When $\bar{\eta}\geq 0.05$, almost no accretion 
takes place.

The numerical results indicate the existence of three-regimes of 
jet formation and accretion depending on the resistivity,
(1) episodic jet formation and accretion for small resistivity: 
($R_{\rm m0}>2.0$),
(2) quasi-steady jet formation and accretion for mildly-resistive
($1<R_{\rm m0}<2.0$) disk and (3) almost no jet formation and no accretion in
highly resistive models ($R_{\rm m0}<1$).
In highly resistive models,
since the torus medium
almost slips the magnetic field lines by magnetic diffusion, magnetic braking 
is insufficient to induce accretion.

Figure 11 shows isocontours of the magnetic Reynolds number 
${\rm log}_{10}R_{\rm m}={\rm log}_{10}(V_{\rm A}\lambda_0 /\eta$). 
Here, $\lambda_0$
is a characteristic scale of the magnetorotational instability 
at $t=0$ at ($r$, $z$) = ($r_0$, $0$), which is defined as 
$\lambda_0=2\pi V_{\rm A0}/\Omega_0$, where $V_{\rm A0}$ and $\Omega_0$ are
the Alfv\'{e}n speed and the rotation angular speed at 
($r$, $z$) = ($r_0$, $0$), respectively. 
In figure 11, the solid curves show the region where
diffusion is not effective ($R_{\rm m} >1$) and
the dotted curves show the region where
diffusion is effective ($R_{\rm m}<1$).
In figure 11a, the diffusive region occupies about half
the volume in the torus. In this model, diffusion does not have 
much affection
on the disk surface where a jet is formed.
No diffusive region appears in figure 11b because model 1 is
a non-diffusive model. 
Figure 11c shows a highly diffusive model.
Since the diffusive region occupies almost the total area in the torus, 
magnetic braking is suppressed and the accretion
is not sufficient
to form a jet.
To illustrate the twist level of 
the magnetic field lines, we show in figure 12 
the time variation of the ratio of 
the toroidal magnetic field, $B_{\varphi}$, to 
the poloidal magnetic field, $B_{\rm p}$,
along a field line determined by integrating 
the equation of the magnetic lines
of force inward from ($r$, $z$) = (1.0, 3.1).
Figure 12 illustrates that the resistive model [model 5; figure 12a, 
model 7; figure 12c] shows smaller twists than that in non-diffusive model
[model 1; figure 12b].
In the non-diffusive model, a strong twist appears at $t=5.5$ and
propagates outward when the jet is created.
Figure 12c illustrates a highly resistive model (model 7) 
which hardly shows outflows 
and has the smallest toroidal magnetic field component.

Figure 13 shows the poloidal velocity, the poloidal fast velocity,
poloidal Alfv\'{e}n velocity, and poloidal slow velocity
along a magnetic field line. 
In model 5 and model 1,
snapshots were taken at 
the start time of jet formation. 
In model 7, the velocities were measured at $t\sim 8\pi$.
Figure 13a (top panels) is for model 5 ($\bar{\eta}=1.25\times 10^{-2}$). 
In model 5, the jet is accelerated
faster than both the Alfv\'{e}n velocity and the slow velocity and 
its speed becomes about 
the Keplerian speed.
Figure 13b (middle panels) is for model 1 ($\bar{\eta}=0$). 
The jet is accelerated similarly to model 5
and the maximum poloidal speed becomes about 1.4 times
the Keplerian speed.
Figure 13c (bottom two panels) is for model 7 
($\bar{\eta}=5.0\times 10^{-2}$).
A jet is not formed and the poloidal velocity does not exceed 
either the slow velocity or the Alfv\'{e}n velocity.

\section
{Discussion}

\subsection{Dependence of the Numerical Results on the Diffusivity}
We have studied the effects of resistivity on the magnetically driven
accretion and jet formation from a torus threaded 
by large-scale poloidal magnetic fields.
When $\bar{\eta}\sim 0$, the surface layer of the torus infalls faster than 
the equatorial region, like an avalanche, due to magnetic braking. 
As the angle between the deformed magnetic field lines and the vertical
direction increases, a magnetically driven jet appears.
The jet formation and mass accretion occur episodically.
This episodic accretion takes place because parts of the torus matter move
radially outward by gaining angular momentum and hinder the mass 
outside of it to accrete.
The magnetorotational instability developing inside the torus
also helps to generate the episodic behavior.
The speed of the jet nearly equals the Keplerian rotation speed.
When $0.0067 < \bar{\eta} < 0.0125$, 
the torus approaches a quasi-steady state without showing episodic
accretion.
When $0.0125<\bar{\eta} <0.05$,  
mass accretion and jet formation
take place, but the mass accretion rate and mass outflow rate decrease 
with $\bar{\eta}$. 
When $\bar{\eta} >0.05$, neither mass accretion nor jet formation takes place.

\subsection{Evaluation of Turbulent Diffusivity}
In this paper, we carried out 2.5-dimensional simulations by explicitly
including uniform magnetic diffusivity.
Even when the classical Spitzer-type resistivity is small,
turbulent diffusivity can be generated by
three-dimensional effects. In three dimensions,
non-axisymmetric instabilities generate magnetic turbulence 
(Hawley et al. 1995; 
Matsumoto, Tajima 1995; Brandenburg et al. 1995). 
The turbulence in accretion disks generates effective magnetic diffusion.
Here, let us estimate the effective diffusivity in an accretion disk 
based on a marginal stability analysis (Matsumoto, Tajima 1995).
The marginal stability model belongs to a self-organized criticality (SOC)
model (e.g., Bak et al. 1988; Mineshige et al. 1994) in complex systems,
which claims that the system spontaneously evolves to a critical state
between unstable states and stable states.
We can obtain a marginally stable state 
by equating the growth rate of the magnetorotational instability, 
$\gamma_{\rm BH}$, and
the stabilization rate due to magnetic diffusivity, $\eta_{\rm t} k^2$,
where $\eta_{\rm t}$ is the turbulent diffusion coefficient and 
$k=2\pi/\lambda_{\rm max}$, here, $\lambda_{\rm max}$ is 
the most unstable wavelength, which we take 
$\lambda_{\rm max}=2\pi V_{\rm A}/\Omega_{\rm K0}$.
In the marginally stable state,
$\gamma_{\rm BH}=\eta_{\rm t} k^2$. 
Since the growth rate of the magnetorotational instability is on the order of
$\gamma_{\rm BH}\sim\Omega_{\rm K0}$,  
\begin{eqnarray}
   \frac{V_{\rm K0} r_0}{\eta_{\rm t}}\sim\frac{V_{\rm K0} r_0}{\Omega_{\rm K0}
   /k^2}
   \sim\frac{(2\pi)^2 V_{\rm K0}r_0}{\Omega_{\rm K0} \lambda_{\rm max}^2} 
   & \nonumber \\
   &\displaystyle
    \mbox{\hspace{-2.33cm}}
    \sim\frac{V_{\rm K0}^2}{V_{\rm A}^2}\sim\frac{\beta}{2}E_{\rm th}^{-1},
\end{eqnarray}
where $\beta =P_{\rm gas}/P_{\rm mag}$. Since $\beta\simeq 10$ in the nonlinear
saturated stage of the magnetorotational instability (Stone et al. 1996;
Matsumoto 1999) and $E_{\rm th}=0.05$ in our model torus, we obtain 
$\eta_{\rm t}/(V_{\rm K0}r_0)\sim 0.01$.
From this result, the effective diffusivity in the marginally stable state 
of the accretion disk can be
estimated as $\bar{\eta}=\eta/(V_{\rm K0}r_0)\sim 0.01$.

We have shown that the numerical results of resistive MHD simulations
sensitively depend on $\bar{\eta}$ when $\bar{\eta}\sim 0.01$.
The order of magnitude estimates of the turbulent resistivity,
$\eta_{\rm t}$ presented here are not sufficient to determine the dependence
of the turbulent resistivity on the disk parameters (e.g., disk thickness).
To obtain exact values of $\eta_{\rm t}$, we need to carry out global
3D MHD simulations.
Hawley (2000) and Machida et al. (2000) published the results of
3D global MHD simulations stating from a geometrically thick disk. 
Much higher resolution simulations are necessary to show the difference
of $\eta_{\rm t}$ for different disk types. Such simulations are now
in progress and will be reported in our subsequent papers.

In this work, we assumed uniform resistivity which mimics the
turbulent diffusion in accretion disks. A rather different model for
dissipation is magnetic reconnection in a hot corona.        
By assuming anomalous resistivity, Hayashi et al. (1996)                 
carried out resistive MHD simulations of magnetic reconnection in magnetic
loops connecting the central star and surrounding accretion disk.
They successfully reproduced the formation of bipolar plasma outflows and
X-ray flares. Magnetic reconnection of magnetic loops whose footpoints
are both on accretion disks have been simulated by Romanova et al. (1998).
The coronal activities of accretion disks acompanying magnetic reconnection
are more complicated processes which cannot be mimiced by
uniform resistivity, and is thus beyond the scope of this paper.

\subsection{Application to AGN Molecular Torus}
Next, we discuss the application of these numerical results to AGNs.
Observations indicate that molecular gas whose mass is 
$10^{10}\mbox{ }M_{\odot}$ 
exists in the central region of high-luminosity IRAS galaxies 
(Scoville et al. 1991) and that a circumnuclear starburst torus
whose radius is between several tens pc and 
1 kpc exists in the central region
of Seyfert galaxies (Wilson et al. 1991, Forbes et al. 1994, 
Storchi-Bergmann et al. 1996).
The ionization rate of the torus may be low because the torus exists far from 
the central black hole. 
We should discuss whether magnetic braking is 
effective in a low-ionized gas torus.
The magnetic diffusivity in partially ionized gas is 
\begin{equation}
   \eta =6.5\times10^{12}({\rm ln}\Lambda)T^{-3/2}
   \left(1+\frac{\tau_{\rm ei}}{\tau_{\rm en}}\right) {\rm cm^2\ s^{-1}}
\end{equation}
and
\begin{equation}
   \frac{\tau_{\rm ei}}{\tau_{\rm en}}=1.3\times 10^{-5}\frac{1}{\chi}
   \left(\frac{T}{\rm 500\ K}\right)^2\frac{1}{\rm ln\ \Lambda}
\end{equation}
(Spitzer 1962). Here, $\tau_{\rm ei}$, $\tau_{\rm en}$, 
$\chi$, and ${\rm ln}\ \Lambda$ are the
electron--ion collision time, electron--neutral atom collision time,
ionization rate,
and Coulomb logarithm ($\sim 10$), respectively.
When the number density of gas satisfies $n\gg 1$,
\begin{equation}
   \eta\simeq8\times 10^3\left(\frac{T}{\rm 500\ K}\right)^{1/2}
   \frac{1}{\chi}
\end{equation}
(Gammie 1996), where the ionization rate, $\chi$, is
\begin{equation}
  \chi=\left\{ \begin{array}{cl}
               10^{-5}n^{-1/2} & 10^4<n<10^{10} \\
               n^{-1}          & 10^{10}<n
              \end{array}
       \right.
\end{equation} 
(Norman, Heyvaerts 1985). 
Assuming that a gas torus exists at 100 pc from the central
black hole and that the temperature of the torus is 100 K, 
the magnetic Reynolds number is
\begin{eqnarray}
   R_{\rm m}\sim\frac{v_{\rm A} H}{\eta}
   \sim 10^{14}\left(\frac{v_{\rm A}}{\rm 1\ km\ s^{-1}}\right)
   \left(\frac{H}{\rm 10\ pc}\right) & \nonumber \\
   & \displaystyle\mbox{\hspace{-3.7cm}}
     \times\left(\frac{T}{\rm 100\ K}\right)^{-1/2}
     \left(\frac{n}{\rm 10^4\ cm^{-3}}\right)^{-1/2}.
\end{eqnarray}
Here, $H$ is the thickness of the gas torus.
Since $R_{\rm m}\gg 1$, magnetic braking is effective 
for charged particles.
We should discuss whether magnetic braking is 
effective for neutral particles, 
too. The velocity of a neutral particle is influenced by ambipolar diffusion.
The velocity difference, $v_{\rm d}$,
between charged particles and neutral particles is
\begin{eqnarray}
   v_{\rm d}\sim
   10^4
   \left(\frac{n}{\rm 10^4\ cm^3}\right)^{-2}
   \left(\frac{\chi}{10^{-7}}\right)^{-1} & \nonumber \\
   & \displaystyle\mbox{\hspace{-3cm}}\times
     \left(\frac{B}{\rm 50\ \mu G}\right)^2
     \left(\frac{H}{\rm 10\ pc}\right)^{-1} {\rm cm\ s^{-1}}
\end{eqnarray}
(Tajima, Shibata 1997).
Since the velocity difference, $v_{\rm d}\sim 10^4{\rm\ cm\ s^{-1}}$, is 
smaller than the dynamical
velocity, $v_{\rm dyna}\sim 10^7{\rm\ cm\ s^{-1}}$,
neutral particles also fall to the black hole 
when magnetic braking takes place around
the surface of the torus.
We conclude that magnetic braking can be effective in 
an AGN gas torus. Recently, Hawley and Stone (1998) carried out 
3D MHD simulations of an ion--neutral fluid.
We would like to confirm the above discussion in the near future 
by extending our MHD code to an ion--neutral MHD code. 

Tables 2 and 3 show the mean and maximum mass outflow rates and accretion rates
when we apply our numerical results to
AGNs which have $M_{\rm BH}=M_{\rm BH9}=10^9\ M_{\odot}$ 
and circumnuclear gas torus with mass $M_{\rm torus}=M_8=10^8\ M_{\odot}$.
In order to explain the energy release rate of quasars, 
\begin{equation}
   L=\varepsilon\dot{M}c^2\sim 3\times 10^{45}
   \left(\frac{\varepsilon}{0.05}\right)
   \left(\frac{\dot{M}}{1\ M_{\odot}\ {\rm yr}^{-1}}\right){\rm erg\ s^{-1}}.
\end{equation}
where $\varepsilon$ is the conversion efficiency of mass energy,
we need a mass accretion rate of at least
$1\ M_{\odot}$ ${\rm yr}^{-1}$.
Table 3 shows that when $\eta <0.05\ V_{\rm K0}r_0$ in an AGN gas torus,
we can explain quasar activity.

\subsection{Order-of-Magnitude Estimation of the Mass Accretion Rate}
We can estimate the mass accretion rate by approximating the torus 
by a rectangular box with height $r_{\rm 0}$. 
The diffusion region of the torus is approximated
by a rectangular box with height $\lambda$, as shown in figure 14. 
The disk gas couples with the magnetic field in a surface region with thickness
$r_{\rm 0}-\lambda$. The accretion rate is
\begin{equation}
   \dot{M}_{\rm acc}=4\pi\int\rho V_{\rm acc} r dz
   \sim 4\pi\times 0.1\rho_0 r_0^2V_{\rm acc}
   \left( 1-\frac{\lambda}{r_0}\right)
   \sim 0.25\rho_0 r_0^3\Omega_{\rm K0}\left( 1-\frac{\lambda}{r_0}\right),
\end{equation}
where
\begin{equation}
   V_{\rm acc}\sim 0.1r_0\Omega_{\rm K0}.
\end{equation}
Here, the density of the torus is assumed to be constant, $\rho=0.25\rho_0$.
The size of $\lambda$ is determined by the condition 
that the magnetic Reynolds number, $R_{\rm m}$, is unity, 
$R_{\rm m}=\lambda V_{\rm A0}/\eta\sim 1$. 
Thus,
\begin{equation}
    \lambda\sim \frac{\eta R_{\rm m}}{V_{\rm A0}}\sim\frac{\eta}{V_{\rm A0}}.
\end{equation}
Using equations (26), (27), and (28), the normalized magnetic diffusivity,
$\bar{\eta}\equiv\eta /(r_0V_{\rm K0}$),
and $E_{\rm mg}=(V_{\rm A0}/V_{\rm K0})^2$ [equation (9)], we can obtain
a normalized mass accretion rate 
\begin{equation}
   \frac{\dot{M}_{\rm acc}}{\rho_0 r_0^2V_{\rm K0}}
   \sim 0.25\left(1-E_{\rm mg}^{-1/2}\bar{\eta}\right).
\end{equation}
This equation can explain the numerically obtained mass accretion rate 
when $0<\bar{\eta}<1.4\times 10^{-2}$,
but can not explain the accretion rate when 
$\bar{\eta}>5.0\times 10^{-2}$
(see figure 15).
To improve this point, we take into consideration the density gradient. 
The mass accretion rate is
\begin{eqnarray}
   \dot{M}\sim \int_{\lambda}^{r_0}4\pi r_0\rho(z) dz V_{\rm acc}
   \sim 4\pi r_0\rho_{\rm surface} HV_{\rm acc} 
   \mbox{\ \ \ \ \ (for $\lambda \ll r_0$)} \nonumber \\
   \sim 10r_0^2\rho_0{\rm\ exp}
   \left[-\frac{1}{2}\left(\frac{\lambda}{H}\right)^2\right]
        H\times 0.1\Omega_{\rm K0} \nonumber \\
   \sim r_0^2\rho_0{\rm\ exp}\left[-\frac{1}{2}
        \left( \frac{\eta \Omega_{\rm K0}}{V_{\rm A0}V_{\rm s0}}\right)^2\right]
        V_{\rm s0} \nonumber \\ 
   \sim \rho_0r_0^2V_{\rm s0}{\rm\ exp}\left[-\frac{1}{2}
        \left( \frac{\bar{\eta}r_0^2\Omega_{\rm K0}^2}{V_{\rm A0}V_{\rm s0}}
        \right)^2\right] \nonumber \\
   \sim \rho_0r_0^2V_{\rm s0}{\rm\ exp}\left[-\frac{1}{2}
        \left( \bar{\eta}E_{\rm mg}^{-1/2}E_{\rm th}^{-1/2}\right)^2\right].
\end{eqnarray}
Here, $E_{\rm th}^{1/2}=V_{\rm s0}/V_{\rm K0}$.
From equation (30), normalized mass accretion rate becomes 
\begin{equation}
   \frac{\dot{M}}{\rho_0r_0^2V_{\rm K0}}
   \sim E_{\rm th}^{1/2}{\rm exp}[-\frac{1}{2}
        \left(\bar{\eta}E_{\rm mg}^{-1/2}E_{\rm th}^{-1/2}\right)^2].
\end{equation}
Here, $H$ is the scale height,
\begin{eqnarray}
   \frac{H}{r_0}\sim\frac{V_{\rm s0}}{V_{\rm K0}}
   \sim \frac{V_{\rm s0}}{r_0\Omega_{\rm K0}}. \nonumber 
\end{eqnarray}
Figure 15 shows the normalized mass accretion rate given 
by equations (29), (31), and the numerical result.
From this figure, the mass accretion rate estimated 
by equation (31)
can better fit the results of numerical simulations.

\subsection{Concluding Remarks}
In this paper, we have presented the results of 2.5-dimensional simulations of
a magnetized torus by explicitly including magnetic diffusivity.
When we carry out three-dimensional simulation, the effects of turbulent
magnetic diffusion automatically comes in when the torus becomes turbulent.
We would like to extend our model to 3D in the near future. 
Furthermore, when we apply the results to AGNs, 
self-gravitational effects
may be important when the torus mass is as large as 
$M_{\rm torus}\sim 10^8\ M_{\odot}$.
We would like to discuss the effects of self-gravity in forthcoming papers.

\vspace{1pc}\par
The numerical computations were carried out on VX/4R at 
the Astronomical Data Analysis Center of 
the National Astronomical Observatory, Japan. 
This work is supported in part by the grant for special field of
Ministry of Education, Sports and Culture (10147105) and Japan Science
and Technology Coorporation (ACT-JST).

\clearpage
\section*
{Reference}
\re
Bak P., Tang C., Wiesenfeld K.\ 1988, Phys.\  Rev.\  Lett.\ A38, 364
\re
Balbus S.A., Hawley J.F.\ 1991, ApJ 376, 214
\re
Blandford R.D., Payne D.G.\ 1982, MNRAS 199, 883
\re
Brandenburg A., Nordlund \AA., Stein R.F., Torkelsson U.\ 1995, ApJ 446, 741
\re
Forbes D.A., Norris R.P., Williger G.M., Smith R.C.\ 1994, AJ 107, 984
\re
Gammie C.F.\ 1996, ApJ 457, 355
\re
Goodson A.P., B\"{o}hm K.-H., Winglee R.M.\ 1999, ApJ 524, 142
\re
Goodson A.P., Winglee R.M., B\"{o}hm K.-H.\ 1997, ApJ 489, 199 
\re
Grosso N., Montmerle T., Feigelson E.D., Andre P., Casanova S.,
Gregorio-Hetem J.\ 1997, Nature 387, 56 
\re
Hawley J.F.\ 2000, ApJ 528, 462
\re
Hawley J.F., Balbus S.A.\ 1992, ApJ 400, 595
\re
Hawley J.F., Gammie C.F., Balbus S.A.\ 1995, ApJ 440, 742
\re
Hawley J.F., Stone J.M.\ 1998, ApJ 501, 758
\re
Hayashi M.R., Shibata K., Matsumoto R.\ 1996, ApJ 468, L37
\re
Hirose S., Uchida Y., Shibata K., Matsumoto R.\ 1997, PASJ 49, 193
\re
Jaffe W., Ford H., Ferrarese L., van den Bosch, F., O'Connell R.W.\ 1996, 
ApJ 460, 214
\re
Kaburaki O.\ 1987, MNRAS 229, 165
\re
Koyama K., Hamaguchi K., Ueno S., Kobayashi N., Feigelson E.D.\ 1996,
PASJ 48, L87
\re
Kudoh T., Matsumoto R., Shibata K.\ 1998, ApJ 508, 186
\re
Kudoh T., Shibata K.\ 1997, ApJ 474, 362
\re
Lovelace R. V. E., Mehanian C., Mobarry C. M., 
Sulkanen M. E.\ 1986, ApJS 62, 1
\re
Lubow S.H., Papaloizou J.C.B., Pringle J.E.\ 1994, MNRAS 268, 1010
\re
Machida M., Hayashi M.R., Matsumoto R.\ 2000, ApJ 532, L67
\re
Matsumoto R.\ 1999, Numerical Astrophysics, eds S.M. Miyama, K. Tomisaka,
T. Hanawa (Kluwer Academic Publishers, Dordrecht), p195
\re
Matsumoto R., Shibata K.\ 1997, Accretion Phenomena and Related Outflows,
IAU Colloquium 163, ASP Conference Series Vol. 121, ed D.T. Wickramasinghe,
G.V. Bicknell, L. Ferrario, p443
\re
Matsumoto R., Tajima T.\ 1995, ApJ 445, 767
\re
Matsumoto R., Uchida Y., Hirose S., Shibata K., Hayashi M.R., Ferrari A.,
Bodo G., Norman C.\ 1996,
ApJ 461, 115
\re
Meier D.L., Edgington S., Godon P., Payne D.G., Lind K.R.\ 1997,
Nature 388, 350
\re
Miller K.A., Stone J.M.\ 1997, ApJ 489, 890
\re
Mineshige S., Ouchi N. B., Nishimori H.\ 1994, PASJ 46, 97 
\re
Norman C., Heyvaerts J.\ 1985, A\&A 147, 247
\re
Ogilvie G.I.\ 1997, MNRAS 288, 63
\re
Ouyed R., Pudritz R.E., Stone J.M.\ 1997, Nature 385, 409
\re
Pudritz R. E., Norman C. A.\ 1983, ApJ 274, 677
\re
Richtmeyer R.D., Morton K.W.\ 1967, Difference Methods for Initial-Value 
Problems, 2nd ed (Interscience Publishers, New York) ch13
\re
Romanova M.M., Ustyugova G.V., Koldoba A.V., Chechetkin V.M.,
Lovelace R.V.E.\ 1998, ApJ 500, 703
\re
Rubin E.L., Burnstein S.Z.\ 1967, J. Comp. Phys. 2, 178
\re
Sano T., Inutsuka S., Miyama S.M.\ 1998, ApJ 506, L57
\re
Scoville N.Z., Sargent A.I., Sanders D.B., Soifer B.T.\ 1991, ApJ 366, L5
\re
Shibata K.\ 1983, PASJ 35, 263
\re
Shibata K., Uchida Y.\ 1986, PASJ 38, 631
\re
Shu F., Najita J., Ostriker E., Wilkin F., Ruden S., Lizano S.\ 1994a,
ApJ 429, 781
\re
Shu F.H., Najita J., Ruden S. P., Lizano S.\ 1994b, ApJ 429, 797
\re
Spitzer L.Jr\ 1962, Physics of Fully Ionized Gases 
(Interscience Publishers, New York), p133 
\re
Stone J.M., Hawley J.F., Gammie C.F., Balbus S.A.\ 1996, ApJ 463, 656
\re
Stone J.M., Norman M.L.\ 1992, ApJS 80, 753
\re
Stone J.M., Norman M.L.\ 1994, ApJ 433, 746
\re
Storchi-Bergmann T., Wilson A.S., Baldwin J.A.\ 1996, ApJ 460, 252
\re
Tajima T., Shibata K.\ 1997, Plasma astrophysics 
(Addison-Wesley, Reading, Massachusetts)
\re
Uchida Y., Shibata K.\ 1985, PASJ 37, 515
\re
Ustyugova G.V., Koldoba A.V., Romanova M.M., Chechetkin V.M., 
Lovelace R.V.E.\ 1995, ApJ 439, L39
\re
Wilson A.S., Helfer T.T., Haniff C.A., Ward M.J.\ 1991, ApJ 381, 79
\re
Yabe T., Aoki T.\ 1991, Comp. Phys. Comm. 66, 219 

\newpage
\begin{table*}[t]
\begin{center}
Table~1.\hspace{4pt}Model ${\rm parameters}^*$.\\
\end{center}
\vspace{6pt}
\begin{tabular*}{\textwidth}{@{\hspace{\tabcolsep}
\extracolsep{\fill}}p{6pc}rc}
\hline\hline\\[-6pt]
Model & $\bar{\eta}=\eta /V_{\rm K0}r_{\rm 0}$ &
$R_{\rm m0}= r_{\rm 0}V_{\rm A0}/\eta$\\[4pt]
\hline\\[-6pt]
1\dotfill & $0$ & $\infty$\\
2\dotfill & $1.25\times 10^{-3}$ & 18\\
3\dotfill & $6.7\times 10^{-3}$ & 3.3\\
4\dotfill & $1.0\times 10^{-2}$ & 2.2\\
5\dotfill & $1.25\times 10^{-2}$ & 1.8\\
6\dotfill & $1.4\times 10^{-2}$ &1.6\\
7\dotfill & $5.0\times 10^{-2}$ &0.4\\[4pt]
\hline
\end{tabular*}
\vspace{6pt}\par\noindent
* In all models,
$E_{\rm th}=V_{\rm s0}^2/(\gamma V_{\rm K0}^2)=0.05$,
$E_{\rm mg}=V_{\rm A0}^2/V_{\rm K0}^2=5\times 10^{-4}$,
and
$\rho_{\rm h}/\rho_0=10^{-3}$.
\end{table*}
\begin{table*}[t]
\begin{center}
Table~2.\hspace{4pt}The mass outflow ${\rm rate}^*$.\\
\end{center}
\vspace{6pt}
\begin{tabular*}{\textwidth}{@{\hspace{\tabcolsep}
\extracolsep{\fill}}p{6pc}rrr}
\hline\hline\\[-6pt]
Model & $\bar{\eta}=\eta/(V_{K0}r_0)$
& $\langle\dot{M}_{\rm jet}/(M_8M_{\rm BH9}^{1/2})\rangle$
& ${\dot{M}_{\rm jet}/(M_8M_{\rm BH9}^{1/2})}_{\rm max}$
\\[4pt]\hline\\[-6pt]
1\dotfill & $0$                   & $5.8$
& $15.5$\\
2\dotfill & $1.25\times 10^{-3}$   & $4.0$
& $18.7$\\
3\dotfill & $6.7\times 10^{-3}$   & $10.8$
& $12.7$\\
4\dotfill & $1.0\times 10^{-2}$   & $13.5$
& $28.4$\\
5\dotfill & $1.25\times 10^{-2}$   & $9.4$
& $14.8$\\
6\dotfill & $1.4\times 10^{-2}$   & $6.1$
& $9.5$\\
7\dotfill & $5.0\times 10^{-2}$   & $0.5$
& $0.9$\\[4pt]
\hline
\end{tabular*}
\vspace{6pt}\par\noindent
* The unit of the mass outflow rate (3rd and 4th columns) is
$M_{\odot}\ {\rm yr}^{-1}$.
\end{table*}
\begin{table*}[t]
\begin{center}
Table~3.\hspace{4pt}The mass accretion ${\rm rate}^*$.\\
\end{center}
\vspace{6pt}
\begin{tabular*}{\textwidth}{@{\hspace{\tabcolsep}
\extracolsep{\fill}}p{6pc}rrr}
\hline\hline\\[-6pt]
Model & $\bar{\eta}=\eta/(V_{K0}r_0)$
& $\langle\dot{M}_{\rm acc}/(M_8M_{\rm BH9}^{1/2})\rangle$
& ${\dot{M}_{\rm acc}/(M_8M_{\rm BH9}^{1/2})}_{\rm max}$
\\[4pt]\hline\\[-6pt]
1\dotfill & $0$
& $26.8$
& $127.9$\\
2\dotfill & $1.25\times 10^{-3}$
& $10.2$
& $38.7$\\
3\dotfill & $6.7\times 10^{-3}$
& $20.8$
& $24.5$\\
4\dotfill & $1.0\times 10^{-2}$
& $20.0$
& $29.0$\\
5\dotfill & $1.25\times 10^{-2}$
& $9.8$
& $16.9$\\
6\dotfill & $1.4\times 10^{-2}$
& $5.3$
& $10.2$\\
7\dotfill & $5.0\times 10^{-2}$
& $0.3$
& $0.5$\\[4pt]
\hline
\end{tabular*}
\vspace{6pt}\par\noindent
* The unit of the mass outflow rate (3rd and 4th columns) is
$M_{\odot}\ {\rm yr}^{-1}$.
\end{table*}

\clearpage
\centerline{Figure Captions}
\bigskip
\begin{fv}{1}
{7cm}
{Schematic picture of the simulation model and 
 simulation box. A differentially rotating gas torus is 
 threaded by global magnetic fields. 
 Numerical simulations were carried out by using 
 an axisymmetric MHD code in cylindrical coordinates.} 
\end{fv}
\begin{fv}{2}
{7cm}
{Time variation of $\rho$ (the contour step width is $0.3$ 
 in logarithmic scale of $\rho$), 
 poloidal velocity vectors {\boldmath $v$}, 
 isocontours of toroidal magnetic field component
 $B_{\varphi}$ (the contour step width is $0.1$), 
 poloidal magnetic field lines $B_{\rm p}$ in model 5 
 ($\bar{\eta}=1.25\times 10^{-2}$).
 The surface layer of the torus falls to the central object 
 because of magnetic braking. 
 A bipolar jet is formed along the large-scale poloidal magnetic field lines
 (see velocity vectors {\boldmath $v$}).}
\end{fv}
\begin{fv}{3}
{7cm}
{Time variation of $\rho$ (the contour step is $0.3$ 
 in logarithmic scale), 
 poloidal velocity vectors {\boldmath $v$}, 
 isocontours of toroidal magnetic field component 
 $B_{\varphi}$ (the contour step is $0.1$), 
 poloidal magnetic field lines $B_{\rm p}$ in model 1 ($\bar{\eta}=0$).
 The magnetic braking is more effective than in model 5 (see $B_{\rm p}$). 
 The magnetic twist is lager than in model 5 (see contours of $B_{\varphi}$) 
 and the jet width is smaller than in model 5 (see {\boldmath $v$}).}
\end{fv}
\begin{fv}{4}
{7cm}
{Time variation of $\rho$ (the contour step is $0.3$ 
 in logarithmic scale), 
 poloidal velocity vectors {\boldmath $v$}, 
 isocontours of toroidal magnetic field component
 $B_{\varphi}$ (the contour step is $0.1$), 
 poloidal magnetic field lines $B_{\rm p}$ 
 in model 7 ($\bar{\eta}=5.0\times 10^{-2}$).
 The magnetic braking is not effective and mass accretion does not 
 take place.
 The magnetic twist is hardly accumulated and a jet is not formed.}
\end{fv}
\begin{fv}{5}
{7cm}
{3D images of the time variation of numerical results for 
 model 5 ($\bar{\eta}=1.25\times 10^{-2}$).
 The top panels show the isosurface of the density and the magnetic field lines.
 The middle panels show the time variation of the two magnetic field lines
        viewed from the side.
 The bottom panels show the time variation of two magnetic field lines
        seen from the top.}
\end{fv}
\begin{fv}{6}
{7cm}
{3D images of the time variation of model 1 ($\bar{\eta}=0$).
 The top panels show the isosurface of the density and the magnetic field lines.
 The middle panels show the time variation of two magnetic field lines
        seen from the side.
 The bottom panels show the time variation of two magnetic field lines
        seen from the top.}
\end{fv}
\begin{fv}{7}
{7cm}
{3D images of the time variation of model 7 ($\bar{\eta}=5.0\times 10^{-2}$).
 The top panels show the isocontour of the density and the magnetic field lines.
 THe middle panels show the time variation of two magnetic field lines
        viewed from the side.
 The bottom panels show the time variation of two magnetic field lines
        seen from the top.}
\end{fv}
\begin{fv}{8}
{7cm}
{Time variation of the temperature distribution (gray scale) and
 $rA_{\varphi}$ ($A_{\varphi}$ is the vector potential),
 which approximately depicts 
 the magnetic field lines and the velocity vectors. 
 The top panels are for model 5. The middle panels are for model 1. 
 The bottom panels are for model 7.}
\end{fv}
\begin{fv}{9}
{7cm}
{Time variation of the mass outflow rate defined as
 $\dot{M}_{\rm jet}=2\int_0^{3.0}2\pi r\rho V_zdr$ at $z=3.1$ for all models.
 As $\bar{\eta}$ increases (from model 1 to model 6), 
 the outflow changes its character from episodic outflow to 
 qusi-steady outflow.
 In a highly diffusive model (model 7), 
 mass outflow hardly takes place.}
\end{fv}
\begin{fv}{10}
{7cm}
{Time variation of the mass accretion rate, defined as
 $\dot{M}_{\rm acc}=2\int_{0}^{0.4}2\pi r\rho V_rdz$ at $r=0.3$ of all models.
 As $\bar{\eta}$ becomes larger (from model 1 to model 6),
 the mass accretion changes from episodic accretion to quasi-steady
 accretion.
 In model 7, mass accretion hardly takes place.} 
\end{fv}
\begin{fv}{11}
{7cm}
{Isocontours of magnetic Reynolds number 
 ($R_{\rm m}=2\pi V_{\rm A} V_{\rm A0}/\eta\Omega_0$) 
 at the initial state ($t= 0$) and 
 at about 2.5 rotation time ($t= 16.4$). (a) model 5 
 ($\bar{\eta}=1.25\times 10^{-2}$),
 (b) model 1 ($\bar{\eta}=0$),
 and (c) model 7 ($\bar{\eta}=5.0\times 10^{-2}$).
 The region depicted by solid curves shows where diffusion is not 
 effective ($R_{\rm m}>1$),
 and the region depicted by dotted curves correspond to
 the diffusive region ($R_{\rm m}<1$).
 In a mildly diffusive model (model 5), 
 the diffusive region occupies about half the area in the torus.
 In the non-diffusive model (model 1), no diffusive region appears.
 In a highly diffusive model (model 7), 
 almost the total region of the torus is diffusive.}
\end{fv}
\begin{fv}{12}
{7cm}
{Ratio of $\mid B_{\rm \varphi}\mid$ to $\mid B_{\rm p}\mid$ 
 along {\boldmath $B$} in (a)  model 5 ($\bar{\eta}=1.25\times 10^{-2}$),
 (b) model 1 ($\bar{\eta}=0$),
 and (c) model 7 ($\bar{\eta}=5.0\times 10^{-2}$).
 In the diffusive model 5 ($\bar{\eta} =1.25\times 10^{-2}$), 
 the twist of magnetic field accumulates gradually. 
 In non-diffusive model 1, 
 magnetic field lines are twisted in short time, and
 we can see that a strong twist propagates along the magnetic field line. 
 In model 7, the twist of magnetic field slowly accumulates,
             but takes long time.  
             Mass outflow and mass accretion 
             hardly occur in this model.}
\end{fv}
\begin{fv}{13}
{7cm}
{Poloidal velocity, poloidal fast velocity, 
 poloidal Alfv\'{e}n velocity and poloidal slow velocity 
 along a magnetic field line 
 for (a) model 5 ($\bar{\eta}=1.25\times 10^{-2}$, $t=16.2$),
 (b) model 1 ($\bar{\eta}=0$, $t=5.8$), and
 (c) model 7 ($\bar{\eta}=5.0\times 10^{-2}$, $t=27$). 
 The filled circles denote the slow point 
 and the open circles show the Alfv\'{e}n point 
 nearest from the equatorial plane.
 }
\end{fv}
\begin{fv}{14}
{7cm}
{Schematic picture of a torus modified 
 to roughly estimate the mass accretion rate. 
 We approximate the cross section of the torus by a square.}
\end{fv}
\begin{fv}{15}
{7cm}
{Comparison between the numerically obtained mass accretion rate
 and analytically obtained
 estimation of the mass accretion rate.
 The equation (31) gives a better fit than equation (29).}
\end{fv}
\end{document}